\documentclass[9pt]{article}
\usepackage{ijcai26}

\usepackage{times}
\usepackage{soul}
\usepackage{url}
\usepackage[hidelinks]{hyperref}
\usepackage[utf8]{inputenc}
\usepackage[small]{caption}
\usepackage{graphicx}
\usepackage{amsmath}
\usepackage{amsthm}
\usepackage{booktabs}
\usepackage{algorithm}
\usepackage{algorithmic}
\usepackage[switch]{lineno}
\usepackage{multirow}

\title{Rigorous Evaluation of Large Language Models for Malaria Drug Discovery: Trade-offs in Performance, Scale, and Resource Utility}

\author{
Marvellous O. Ajala$^1$
\and
Zainab Ashimiyu-Abdusalam$^1$\and
Comfort Adesina$^1$\\
\affiliations
$^1$Magami Open Sciences Initiative\\
\emails
\{marvellous, zainab, comfort\}@magamios.org
}

\begin{document}

\maketitle

\begin{abstract}
We introduce Malaria-Instruct, a curated instruction-following dataset derived from the ChEMBL Legacy Malaria corpus for Malaria virtual screening, and conduct a systematic evaluation of five open-source LLMs; Gemma-2 2B/9B, TxGemma-2B/9B, and LlaSMol-Mistral-7B, on a rigorous out-of-distribution data split. Performance was benchmarked against classical ML models (Random Forest, XGBoost) and frontier proprietary models (Gemini 2.5, OpenAI o3) under few-shot conditions. Fine-tuned LLMs substantially outperformed all baselines: TxGemma-9B achieved the highest ROC-AUC (0.731 ± 0.005) and LlaSMol-Mistral-7B the best enrichment factor (EF@1\% $\approx$ 4.99). Domain-specific fine-tuning proved categorically indispensable with TxGemma-9B collapsing from ROC-AUC 0.731 to 0.499, under its best few-shot condition, and neither Gemini 2.5 (ROC-AUC $\approx$ 0.53) nor o3 (ROC-AUC $\approx$ 0.59) achieved reliable discrimination without fine-tuning. Biomedical pretraining conferred a measurable advantage at equivalent scale, while chemistry-aware pretraining yielded superior prospective enrichment. Fine-tuned open-source LLMs represent a compelling, resource-efficient paradigm for antimalarial VS, outperforming both classical pipelines and proprietary reasoning models under structurally challenging conditions.
\end{abstract}

\section{Introduction}
\subsection{The Global Burden of Malaria and Advent of Virtual Screening}\label{the-global-burden-of-malaria-and-the-urgency-for-novel-therapeutics}
Malaria remains one of the most consequential infectious diseases confronting global public health, with an estimated 263 million cases and 597,000 deaths recorded globally in 2023 ~\cite{who2024malaria} and approximately 94\% of all cases and 95\% of deaths occurred in the WHO African Region. These figures underscore the profound human cost of the disease and the persistent structural inequities that concentrate its burden in low-resource settings.
The emergence of partial artemisinin resistance (the current frontline standard of care following the progressive failure of chloroquine and sulfadoxine--pyrimethamine) in Africa ~\cite{who2024malaria,lancet2025malaria}, following its spread from Southeast Asia, urgently motivates the identification of structurally novel chemotypes.
Ligand-based virtual screening computationally ranks candidate molecules by predicted bioactivity, evaluated here primarily by enrichment factor at 1\% (EF@1\%), the operationally critical metric governing hit rate in experimental screening. For antimalarials specifically, the central challenge is that effective VS must generalise to structurally novel chemotypes beyond the historically narrow scaffold space of known actives, precisely the out-of-distribution condition this study is designed to test.

\subsection{The Rise of Large Language Models in Molecular Biology and Cheminformatics}\label{the-rise-of-large-language-models-in-molecular-biology-and-cheminformatics}
The representation of molecular structures as SMILES strings \cite{weininger1988smiles},that are amenable to sequence modelling, catalysed a proliferation of chemistry-aware LLMs \cite{chithrananda2020chemberta,fabian2020molecular,yu2024llasmol}. More recent developments, including the SMolInstruct-trained LlaSMol family \cite{yu2024llasmol}, the therapeutics-focused TxGemma series \cite{wang2025txgemma}, and the predecessor Tx-LLM \cite{zambrano2024txllm}, have produced instruction-following models that accept molecular queries in natural language, respond with predicted properties, and exhibit strong transfer across diverse chemical tasks. 
The emergence of frontier proprietary LLMs with broad reasoning capabilities, notably OpenAI\textquotesingle s o3 and Google\textquotesingle s Gemini 2.5, raises a further question of practical significance: can general-purpose reasoning models leverage their extensive scientific pretraining to perform useful bioactivity prediction in a few-shot setting, without domain-specific fine-tuning? 

\subsection{Fine-Tuning vs. Few-Shot Prompting: A Critical Distinction for Specialised Scientific Tasks}\label{fine-tuning-vs.-few-shot-prompting-a-critical-distinction-for-specialised-scientific-tasks}
A foundational distinction in the deployment of LLMs for specialised scientific tasks concerns the mechanism by which task-specific knowledge is conveyed to the model.
Parameter-efficient fine-tuning (QLoRA) internalises structure-activity relationships at the gradient level in a way that ICL cannot \cite{hu2021lora,brown2020language}, making domain-specific adaptation a testable necessity rather than a stylistic choice.
Bioactivity prediction is not primarily a reasoning task amenable to pattern retrieval from a few examples as it requires the model to encode fine-grained structure-activity relationships that may not be recoverable from the statistical regularities present in even a handful of SMILES-label pairs. The hypothesis that domain-specific fine-tuning is necessary for reliable bioactivity prediction, rather than optional, is a central empirical claim of this work.

\subsection{The Generalisation Problem: Why Scaffold Dissimilarity Splitting Matters}\label{the-generalisation-problem-why-scaffold-dissimilarity-splitting-matters}
A recent study by \cite{meidi2024dataset} indicates that many splitting techniques provides negligible discriminative properties and that only dissimilarity-based and clustering-based splitting methods provide a meaningful test of out-of-distribution generalisation.
In this study, we adopt the Lo-Hi dissimilarity splitting framework as the partitioning standard, enforcing strong Tanimoto dissimilarity train--test and train--validation, applied independently per assay. This design choice is deliberate and consequential: it ensures that reported performance metrics reflect true generalisation capacity rather than scaffold memorisation, and provides a substantially more demanding and ecologically valid benchmark than random or scaffold splitting would permit.

\subsection{Research Contribution}\label{gaps-in-the-literature-and-research-objectives}
Our primary contributions in this work are: (1) We introduce \textbf{Malaria-Instruct}, a novel instruction-following dataset derived from the ChEMBL Legacy Malaria corpus, curated with rigorous deduplication, assay harmonisation, and expert-informed contextualisation. (2) We establish a rigorous benchmark using Lo-Hi dissimilarity-based splitting that enforces meaningful out-of-distribution generalisation conditions, providing systematic head-to-head comparison of fine-tuned open-source LLMs against frontier proprietary models and classical cheminformatics baselines where non previously existed. (3) We provide a comprehensive evaluation of five open-source LLMs across fine-tuning and few-shot paradigms, with systematic characterisation of resource requirements to guide deployment decisions in low-resource research settings prevalent in Africa where these researches are often undertaken.

\section{Related Works}\label{related-works}
\subsection{ChEMBL and Curated Antimalarial Bioactivity Databases}\label{chembl-and-curated-antimalarial-bioactivity-databases}
Malaria-Instruct is built on ChEMBL \cite{mendez2019chembl,zdrazil2024chembl}, (details provided in appendix A)  while addressing the while addressing the key curation challenges that exist in it: class imbalance, assay heterogeneity (across different organisms, time points, and measurement modalities), and duplicate bioactivity records (from the same molecule being tested across multiple experiments) persists. We address these challenges through a multi-stage curation procedure involving assay-level deduplication, harmonisation of 48-hour and 96-hour readouts, and ECFP-based negative sample augmentation for confirmatory assays lacking sufficient inactive observations.
\subsection{Instruction Tuning and Dataset Curation for Scientific LLMs}\label{instruction-tuning-and-dataset-curation-for-scientific-llms}
Unlike prior general-purpose molecular instruction datasets \cite{fang2023molinstructions,yu2024llasmol,cao2024instructmol} discussed in appendix B, Malaria-Instruct is the first instruction-tuning dataset specifically constructed for antimalarial virtual screening, with assay-level contextualisation providing information about Plasmodium strain, assay duration, and mechanistic context. This contextualisation is specifically designed to support TxGemma\textquotesingle s \cite{wang2025txgemma} prompt format, which conditions predictions on additional biological context beyond the molecular structure alone. The dataset\textquotesingle s construction incorporates per-assay, per-split negative sample augmentation.
\subsection{Chemistry-Aware and Biomedically Specialised Language Models}\label{chemistry-aware-and-biomedically-specialised-language-models}
The application of language modelling to molecular SMILES strings has a well-established lineage including ChemBERTa ~\cite{chithrananda2020chemberta} and MolBERT ~\cite{fabian2020molecular} 
The most directly relevant prior work to this study is LlaSMol models ~\cite{yu2024llasmol}, which were produced by fine-tuning four open-source base models (Galactica, Llama 2, Code Llama, and Mistral) on SMolInstruct using LoRA. The Mistral-based variant, LlaSMol-Mistral, was identified as the best-performing chemistry LLM, outperforming GPT-4 and Claude Opus 3 by substantial margins on standard chemistry benchmarks at the time of publication. LlaSMol\textquotesingle s chemistry-specific pretraining provides a principled inductive bias for molecular property prediction that general-purpose models lack, making it a particularly informative comparator in our evaluation.
Domain-specific pretraining in therapeutic AI, grounded in the recognition that biomedical language: spanning clinical ontologies, protein nomenclature, molecular property distributions, and assay-specific terminology, constitutes a distinct subdomain that general-purpose models underrepresent relative to its downstream task density. TxGemma\cite{wang2025txgemma} extends Tx-LLM ~\cite{zambrano2024txllm} achieving near-state-of-the-art performance on 43 of 66 Therapeutics Data Commons (TDC) tasks ~\cite{huang2021tdc}, details provided in appendix C. The biomedical specialisation of TxGemma makes it a compelling candidate for antimalarial bioactivity prediction, but the extent to which its pretraining generalises to the specific structural and biological features of the Plasmodium screening context remains an open empirical question.
While general-purpose LLMs like GPT-4 and the reasoning-focused o-series show promise in diverse scientific tasks, they consistently underperform compared to specialized models in predicting molecular bioactivity. This suggests that the "reasoning-by-analogy" used by these models cannot capture the high-dimensional, idiosyncratic structural patterns required for accurate drug-activity interpolation.

\subsection{Data Splitting Strategies \& Evaluation Frameworks in Cheminformatics}\label{data-splitting-strategies-evaluation-frameworks-in-cheminformatics-benchmarking}
The choice of dataset splitting methodology is among the most
consequential methodological decisions in molecular ML benchmarking, yet it has received insufficient attention in many published evaluations. Random splitting, the default in general ML frameworks, produces inflated performance estimates for molecular models due to the high structural similarity of molecules within the same dataset, effectively allowing models to interpolate between highly similar training and test compounds. Scaffold splitting \cite{bemis1996properties} provides a more structurally aware partition but does not explicitly control the degree of inter-partition Tanimoto similarity, leaving open the possibility of near-analogues spanning the split boundary.
The Lo-Hi framework \cite{steshin2023lohi} operationalises dissimilarity-based splitting as a graph-theoretic optimisation problem: constructing a maximum independent set partition such that no two molecules from different partitions share a Tanimoto similarity above the specified threshold.
Meidi \textit{et al.,} (2024) \nocite{meidi2024dataset} provided a systematic comparison of splitting strategies across multiple bioactivity benchmarks, demonstrating that only dissimilarity-based and clustering-based splits (group structurally similar compounds and assign whole clusters to the same partition) provide a reliable test of out-of-distribution generalisation, with random and scaffold splits substantially overestimating practical model performance. Thus, the Lo-Hi splitting framework is adopted in this study as the standard for all partitioning operations, applied independently per assay to respect the distinct chemical space coverage of each screening dataset.
Also, the choice of evaluation metric profoundly shapes the conclusions drawn from VS model benchmarks. In appendix D, we provide an extensive discussion on the different considerations for evaluation metrics. However, from the perspective of prospective VS deployment, we believe the Enrichment Factor (EF) is arguably the most operationally relevant metric. EF@1\% quantifies the fold-enrichment of true actives in the top 1\% of model-ranked compounds relative to the expected random baseline, directly measuring the practical efficiency of a VS campaign. A model with EF@1\% of 5.0 concentrates five times as many true actives in the top 1\% of its predictions as would be expected by chance, a difference that translates directly into reduced experimental costs and accelerated hit identification. The tension between ROC-AUC optimisation and EF optimisation, which arise from different parts of the score distribution, is a substantive methodological issue explored in the Discussion.

\section{Methodology}\label{methodology}
\subsection{Dataset}\label{methodology-dataset}
\subsubsection{Dataset Curation}\label{methodology-dataset-curation}
We introduce \textbf{Malaria-Instruct}, a novel instruction-following dataset for molecular bioactivity prediction constructed from the ChEMBL Legacy Malaria corpus \cite{mendez2019chembl,zdrazil2024chembl}. \footnote{The Malaria-Instruct dataset is publicly available on Zenodo at \url{https://zenodo.org/records/19222923}; the evaluation and fine-tuning code is available on GitHub at \url{https://github.com/Magami-Open-Sciences-Initiative/LLMS-for-Malaria}.}

\emph{Assay selection and deduplication.} Only potency and IC50 assay entries were retained; all other bioactivity modalities were excluded. Because biological assays are routinely conducted in technical replicates, duplicate records were resolved at the assay level: for each unique molecule--assay pair, all replicate readings were aggregated and the datapoint was assigned a positive label if all replicates were concordantly active, a negative label if all replicates were concordantly inactive, and removed entirely if replicates returned conflicting classifications. This conservative conflict-removal policy prioritises label precision over dataset size, accepting a reduction in the number of usable datapoints to avoid the introduction of ambiguous supervision signals.

\emph{Temporal harmonisation.} Antimalarial assays are standardly designed for 48-hour or 96-hour incubation periods, with intermediate 24-hour readings sometimes recorded. For assays with 24-, 48-, and 96-hour readings, labels were harmonised using the same concordance rule; conflicting timepoint readings were discarded.

\emph{Assay contextualisation.} TxGemma\textquotesingle s prompt format conditions predictions on structured biological context beyond the molecular SMILES string alone, including information about assay type, biological target, and experimental conditions \cite{zambrano2024txllm} \cite{wang2025txgemma}. To support this, an additional contextualisation column was constructed for each assay entry, encoding the nature of the assay, the specific \emph{Plasmodium} strain under investigation, and distinguishing experimental characteristics. This contextualisation was produced through a combination of domain expert annotation and structured elicitation from state-of-the-art LLMs operating under expert-validated templates, ensuring biological accuracy while maintaining scalability across the full assay inventory. \footnote{This contextualisation was not added to the prompts for LlaSmol sticking with the formatting style used to train the model, leveraging only assay level details}.

\emph{Negative sample augmentation.} Confirmatory assays, designed as follow-up screens to validate hits from primary HTS campaigns, contain exclusively or predominantly active compounds by construction, providing no inactive observations from which a decision boundary can be learned. For such assays, negative sample augmentation was performed on a per-assay, per-split after the train-test-validation partitioning. ECFP4 fingerprints (radius 2, 2048 bits) were computed for molecules from previously discarded assays and dimensionality-reduced via PCA to the most informative components, retaining those explaining the greatest variance subject to a maximum of 100 dimensions or the number of available samples, whichever was smaller. The resulting embeddings were further projected to two dimensions via t-SNE using the Euclidean metric. Candidate negatives were then selected from the spatial region circumscribed by the convex envelope of confirmed positive samples in the t-SNE embedding, ensuring that augmented negatives occupy chemically plausible proximity to the active scaffold space rather than being drawn from remote or trivially dissimilar regions of chemical space. This approach was taken as it currently encapsulates how chemists currently computationally explore chemical space. Augmentation was performed to restore the positive-to-negative ratio observed in the broader Malaria-Instruct corpus. \footnote{Augmented negatives were constrained to the pharmacophoric neighbourhood of training actives to ensure decision-boundary relevance.}

\emph{Molecular standardisation.} All retained molecules were standardised using the following sequential operations: charge neutralisation, parent fragment extraction for multi-component mixtures, normalisation of unusual valence states and functional group representations, and canonical tautomer selection. Standardisation was applied uniformly across all splits to ensure that molecular identity comparisons and fingerprint computations reflect a consistent chemical representation.

\subsubsection{Data Splitting}\label{methodology-data-splitting}
Lo-Hi splitting \cite{steshin2023lohi} was adopted following \cite{meidi2024dataset}, who demonstrate only dissimilarity-based splits provide genuine out-of-distribution evaluation.
Partitioning was performed on ECFP4 fingerprints (radius 2, 2048 bits), selected for maximum substructural expressivity. Pairwise Tanimoto similarities were computed across the full molecular inventory, and partitions were constructed such that the maximum Tanimoto similarity between any train--test molecule pair did not exceed 0.4, and the maximum similarity between any train--validation pair did not exceed 0.55. These thresholds enforce a structural dissimilarity condition that approximates the novelty conditions encountered in real-world hit identification campaigns. Splitting was performed independently for each assay, preserving the structural diversity constraints within each biological context rather than applying a global partition that could allow assay-level leakage. For fine-tuned model experiments, an independent Lo-Hi resplit was drawn for each replicate, ensuring that performance standard deviations reflect genuine partition-induced variance rather than repeated evaluation on an identical structural partition.

\subsubsection{Data Preparation}\label{methodology-data-preparation}
\emph{Fine-tuning format.} Training instances for all fine-tuned LLMs were formatted following the data preparation protocol introduced by Tx-LLM \cite{zambrano2024txllm}. The resulting training corpus contains a mixture of 70\% few-shot-formatted examples, with the number of in-context examples per instance drawn uniformly from \{2, 3, 4, 5\} and 30\% zero-shot examples. For few-shot training instances, both the in-context demonstration molecules and the query molecule for which a prediction is requested were drawn exclusively from the training partition, ensuring no validation or test information was accessible during training. Test and validation sets were formatted exclusively as zero-shot instances for final evaluation.

\emph{Few-shot evaluation format.} For few-shot evaluation of unfinetuned models, in-context demonstration examples were drawn from either the training or test partitions, while the query molecule for which a prediction is requested was drawn strictly from the validation set. For each shot-count condition (3-shot, 4-shot, 5-shot), the dataset was independently resplit prior to constructing the few-shot evaluation instances, such that each shot-count condition operates on a distinct structural partition. The inferential consequences of this design for variance interpretation are discussed in §4.6.

\subsection{Modelling}\label{methodology-modelling}
Model evaluation was structured around two complementary paradigms: supervised fine-tuning and few-shot in-context learning, applied to model classes spanning classical machine learning baselines, general-purpose open-source LLMs, and domain-specialised open-source LLMs, with additional few-shot evaluation of frontier closed-source models.
\subsubsection{Fine-Tuning}\label{methodology-finetuning}
\emph{Classical baselines.} Random Forest \cite{breiman2001random} and XGBoost \cite{chen2016xgboost} were trained on 2048-bit Morgan fingerprints with radius 2, computed as binary numpy arrays for each molecule. Both models were trained with default hyperparameter configurations and a fixed random state of 2024 and experiments conducted across five independent replicates with resplitting at each replicate. Justification for the choice of baseline provided in appendix E.

\emph{LLM fine-tuning.} Three open-source model families were selected for parameter-efficient fine-tuning: Gemma-2 \cite{riviere2024gemma2}, TxGemma \cite{wang2025txgemma}, and LlaSMol-Mistral \cite{yu2024llasmol}. Gemma-2 was included as a general-purpose reference to quantify the performance gap attributable to domain-specific pretraining. TxGemma represents the current state of the art among open-source therapeutic LLMs, built on the Gemma-2 architecture and pre-trained on an instruction-tuned variant of the TDC benchmark \cite{huang2021tdc}. LlaSMol-Mistral, based on the Mistral-7B backbone, was selected as the best-performing variant from the LlaSMol model family, itself fine-tuned on the SMolInstruct dataset \cite{yu2024llasmol}. Both 2B and 9B parameter variants were evaluated for Gemma-2 and TxGemma to characterise the performance--resource trade-off across the small-to-medium scale range most relevant to resource-constrained research environments. Training parameters are defined in appendix F.
All LLM experiments were conducted in duplicate with independent Lo-Hi resplitting at each replicate. Training used the training partition; evaluation during training monitored the test partition. Final performance metrics are reported exclusively on the validation set. Validation predictions were generated via vLLM to reduce inference latency.

\emph{Compute environment.} All training was conducted on Google Colab Pro+ (paid tier); compute requirements are reported in Table 4
\subsection{Few-Shot Evaluation}
\emph{Open-source models.} The in-context learning capability of all unfinetuned open-source models was evaluated under 3-shot, 4-shot, and 5-shot conditions. At each shot count, \$n\$ molecule--activity demonstration pairs were prepended to the query prompt prior to soliciting a prediction for the target molecule. See §4.6 and appendix G for inferential implications of per conditioning splitting.

\emph{Closed-source models.} Gemini 2.5 and OpenAI o3 were evaluated under identical 3-, 4-, and 5-shot conditions via their respective public APIs. Fine-tuning was not pursued for closed-source models, as API-based fine-tuning is inconsistent with the resource-constrained, open-infrastructure research environment this study targets. To emulate the low-resource constraint predominant with researchers carrying out similar research in global south associated with API token costs, evaluation was conducted on a subsample of 500 molecules per condition\footnote{a sample size was chosen to exceed the minimum of 326 derived via Finite Population Correction at 95\% confidence and 5\% margin of error, providing statistically adequate coverage of the validation set distribution}. All closed-source evaluations were conducted in duplicate at each shot-count condition, with mean and standard deviation reported across replicates.

\subsubsection{Evaluation}\label{methodology-evaluation}
All models were evaluated on the held-out validation set, which was retained as zero-shot examples across both fine-tuned and few-shot experimental conditions, ensuring that no validation molecule or its structural neighbours appeared in any prompt context seen during training or few-shot construction. Model outputs were parsed to extract binary activity predictions, and probabilistic scores were derived from output token logits where available, or from the rank ordering of predicted class labels otherwise.
Performance was assessed across five complementary metrics selected to capture distinct and non-redundant aspects of model behaviour under class imbalance. Accuracy and precision are reported for completeness but are not treated as primary metrics, given their susceptibility to inflation by majority-class prediction under the positive-to-negative imbalance characteristic of antimalarial screening data. The Matthews Correlation Coefficient (MCC) serves as the primary balanced classification metric: unlike F1, MCC incorporates all four cells of the confusion matrix and remains informative under severe class imbalance, producing a score of zero for a classifier that predicts the majority class unconditionally \cite{chicco2020advantages}. ROC-AUC is reported as the standard global discrimination metric, corresponding to the probability that a randomly drawn active compound is ranked above a randomly drawn inactive, and is insensitive to the choice of classification threshold.
The Enrichment Factor at the 1\% level (EF@1\%) is designated the primary operational metric for virtual screening evaluation. EF@1\% quantifies the fold-enrichment of confirmed actives within the top 1\% of model-ranked compounds relative to the expected rate under random selection, and directly indexes the practical efficiency of a prospective screening campaign, governing the number of experimental assays required per confirmed hit. A model achieving EF@1\% of 5.0 concentrates five times the expected number of true actives in the top-ranked fraction, reducing experimental cost per hit by the same factor. Where ROC-AUC and EF@1\% rankings diverge across models, EF@1\% is treated as the operationally authoritative criterion for model selection.
For fine-tuned models, all metrics are reported as mean ± standard deviation across the two independent experimental replicates. For few-shot open-source models, metrics are reported per shot count across the independently resplit evaluation conditions. For closed-source models, evaluated on a statistically powered subsample of 500 molecules, exceeding the minimum sample size of 326 derived via Finite Population Correction at 95\% confidence and 5\% margin of error, metrics are reported as mean ± standard deviation across duplicate runs at each shot count. Inference time per molecule and GPU memory requirements during both training and inference are additionally recorded for all models to support resource-constrained deployment decisions, and are summarised alongside performance metrics in Table 1.

\section{Results} \label{results}
We provide the mean value of the results for the trained classical models, finetuned LLMs and the best value for the ICL of the closed source LLMs and the best values highlighted. TxGemma 9B gave the best ROC\_AUC value (0.7315 $\pm$ 0.0053) while LlaSmol-Mistral had the best EF1\% (4.9865 $\pm$ 0.0048) and MCC (0.56415 $\pm$ 0.0257) while XGBoost gave the best inference time of 0.00004s  per sample. In Table 2, we present the result for the few-shot learning of the open source models. Except LlaSmol with accuracies within the range of 0.7952 \- 0.8020 for 3 \- 5 shots, all other models have accuracies less than 0.3 (Gemma 9B), 0.08 (TxGemma 2B) and 0.008 (TxGemma 9B). LlaSmol also had the highest MCC (0.0721 at 4 shots) while Gemma 9B had the highest AUC (0.5845 at 3 shots) and TxGemma had the highest EF1\% (1.8160 at 3 shots). In Table 3, we present the result (mean) for the few-shot learning of the closed source models, with OpenAI o3 giving the higher AUC values (highest - 4 shots 0.5907) while Gemini 2.5 had the higher EF1\% (highest - 4 shots, 3.788). For the MCC value, while o3 on average had higher values compared to Gemini, Gemini at 4 shots had the highest MCC (0.1689). In table 4, we present the compute requirement for training and inference. During training, the 2B models were trained on 16GB GPU while for inference, all models required atleast 24GB GPU for inference. Also, Gemma 2B had the best inference time (s) per molecule of 0.004s after finetuning while TxGemma 2B had the best inference time (s) per sample (0.0142s) during few-shoting. This value is further explored in Figure 1 and 2 where the inference time (s) per molecule is plotted against EF1\% and AUC.

\begin{table*}[t]
\centering
\small
\caption{General Performance results of the experiments.}
\resizebox{\textwidth}{!}{
\begin{tabular}{lllccccr}
\toprule
\textbf{Category} & \textbf{Task Type} & \textbf{Model} & \textbf{Size} & \textbf{ROC\_AUC} & \textbf{MCC} & \textbf{EF (1\%)} & \textbf{Inf. Time (s)} \\
\midrule
Classical & Finetuned & RandomForest & -- & 0.6828  & 0.2581  & 2.5790  & 0.0004  \\
& Finetuned & XGBoost & -- & 0.6652  & 0.2164  & 1.7908  & \textbf{0.00004}  \\
\midrule
Open Source & Finetuned & Gemma 2 & 2B & 0.7063  & 0.47735  & 3.66045  & 0.0040  \\
LLMs & Finetuned & TxGemma 2 & 2B & 0.6955  & 0.53275  & 4.6646  & 0.0047  \\
& Finetuned & LlaSmol - Mistral & 7B & 0.70225  & \textbf{0.56415}  & \textbf{4.9865}  & 0.08015  \\
& Finetuned & Gemma 2 & 9B & 0.6383  & 0.4408  & 4.6347  & 0.0304  \\
& Finetuned & TxGemma 2 & 9B & \textbf{0.73155}  & 0.5539  & 4.23025  & 0.03705  \\
\midrule
Closed Source & Fewshot & OpenAI-o3 & -- & 0.5284  & 0.1689  & 3.7879  & -- \\
LLM & (Best) & Gemini 2.5 & -- & 0.5907  & 0.1584  & 1.64635  & -- \\
\bottomrule
\end{tabular}}
\end{table*}

\begin{table}[h]
\centering
\footnotesize
\caption{Evaluation Metrics of the Fewshot Open Source Models}
\begin{tabular}{lccccccc}
\toprule
\textbf{Model} & \textbf{Shots} & \textbf{Acc} & \textbf{AUC} & \textbf{MCC} & \textbf{EF} \\
\midrule
\multirow{3}{*}{LlaSMol 7B} & 3 & 0.7952 & 0.5140 & 0.0545 & 1.5424 \\
& 4 & 0.8011 & 0.5196 & \textbf{0.0721} & 1.6980 \\
& 5 & \textbf{0.8020} & 0.5128 & 0.0455 & 1.4416 \\
\midrule
\multirow{3}{*}{Gemma 2B} & 3 & 0.0061 & 0.5025 & -0.0026 & 0.2971 \\
& 4 & 0.0059 & 0.5029 & 0.0053 & 0.7198 \\
& 5 & 0.0009 & 0.4993 & 0.0038 & 1.2486 \\
\midrule
\multirow{3}{*}{TxGemma 2B} & 3 & 0.075 & 0.5363 & -0.0258 & 0.8036 \\
& 4 & 0.0698 & 0.5268 & -0.0221 & 0.8042 \\
& 5 & 0.0915 & 0.5535 & -0.0156 & 0.669 \\
\midrule
\multirow{3}{*}{Gemma 9B} & 3 & 0.2930 & \textbf{0.5845} & 0.0653 & 1.1100 \\
& 4 & 0.2622 & 0.5571 & 0.0475 & 1.1095 \\
& 5 & 0.2850 & 0.5697 & 0.0525 & 1.0761 \\
\midrule
\multirow{3}{*}{TxGemma 9B} & 3 & 0.0073 & 0.4892 & 0.0290 & \textbf{1.8160} \\
& 4 & 0.0028 & 0.4991 & 0.0029 & 1.1037 \\
& 5 & 0.0023 & 0.4984 & 0.0059 & 1.2532 \\
\bottomrule
\end{tabular}
\end{table}

\begin{table}[h]
\centering
\footnotesize
\caption{Evaluation Metrics of the Fewshot Closed Source Models[cite: 39].}
\begin{tabular}{lccccc}
\toprule
\textbf{Model} & \textbf{Shots} & \textbf{ROC\_AUC} & \textbf{MCC} & \textbf{EF (1\%)} \\
\midrule
\multirow{3}{*}{Gemini 2.5} & 3 & 0.5182 & 0.0945 & 2.342 \\
& 4 & 0.5284 & \textbf{0.1689} & \textbf{3.788} \\
& 5 & 0.5175 & 0.0883 & 2.254 \\
\midrule
\multirow{3}{*}{OpenAI o3} & 3 & 0.5751 & 0.1285 & 1.411 \\
& 4 & \textbf{0.5907} & 0.1573 & 1.551 \\
& 5 & 0.5867 & 0.1583 & 1.646 \\
\bottomrule
\end{tabular}
\end{table}

\begin{table}[h]
\centering
\footnotesize
\caption{Compute requirement table of the finetuning and fewshoting task[cite: 101].}
\resizebox{\columnwidth}{!}{
\begin{tabular}{llccc|ccc}
\toprule
& & \multicolumn{3}{c}{\textbf{Finetuning}} & \multicolumn{3}{c}{\textbf{Fewshot (Best)}} \\
\textbf{Size} & \textbf{Model} & \textbf{Train GPU} & \textbf{Inf. Time} & \textbf{EF} & \textbf{Inf. GPU} & \textbf{Inf. Time} & \textbf{EF} \\
\midrule
2B & Gemma 2 & 16GB & 0.0040 & 3.660 & 24GB & 0.0457 & 0.720 \\
2B & TxGemma & 16GB & 0.0047 & 4.665 & 24GB & 0.0142 & 0.804 \\
7B & LlaSmol & 24GB & 0.0802 & 4.987 & 24GB & 0.2337 & 1.698 \\
9B & Gemma 2 & 24GB & 0.0304 & 4.635 & 24GB & 0.4895 & 1.110 \\
9B & TxGemma & 24GB & 0.0371 & 4.230 & 24GB & 0.0873 & 1.816 \\
\bottomrule
\end{tabular}}
\end{table}

\begin{figure}[h]               
  \centering                  
\includegraphics[width=3.20in]{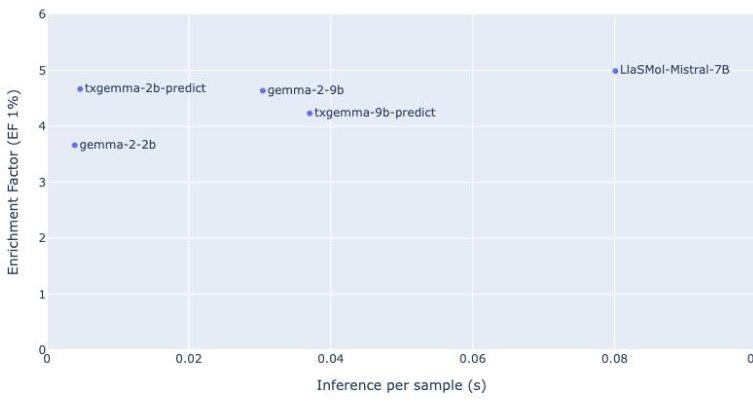}
  \caption{Plot of the Enrichment Factor 1\% against inference time (s) per molecule.}
\end{figure}

\begin{figure}[h]               
  \centering                  
\includegraphics[width=3.20in]{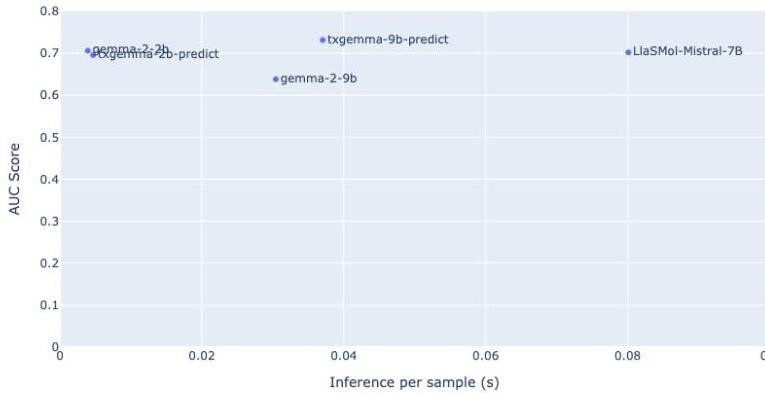}
  \caption{Plot of the AUC Score against inference time (s) per molecule } 
  \label{fig:my_unique_label}             
\end{figure}

\section{Discussion}\label{discussion}
The results of this study yield several interconnected findings that collectively illuminate the conditions under which LLMs can contribute meaningfully to antimalarial virtual screening, the boundaries of their utility, and the trade-offs between performance, domain specialisation, and computational resource requirements. We discuss these findings in turn, situating them within the broader context of molecular ML methodology.
\subsection{Fine-Tuning is Non-Negotiable: The Collapse of Few-Shot Performance Across All Model Classes}\label{fine-tuning-is-non-negotiable-the-collapse-of-few-shot-performance-across-all-model-classes}
The most consequential finding of this study is the universal failure of few-shot in-context learning for antimalarial bioactivity prediction, observed across all model classes, including frontier proprietary models with reasoning capabilities. Notably, TxGemma-9B, which achieves the highest ROC-AUC (0.731) among fine-tuned models, collapses to ROC-AUC $\approx$ 0.499 under 4-shot prompting (as well as $<$0.01 accuracy), representing essentially random discrimination. This dramatic inversion cannot be attributed to insufficient model capacity or general reasoning ability: the same model, when fine-tuned with gradient-level domain adaptation, achieves strong discriminative performance.
The performance of closed-source frontier models further underscores this conclusion. Gemini 2.5, evaluated under 3-shot conditions on a statistically powered subsample of 500 molecules, achieves ROC-AUC $\approx$ 0.52 indistinguishable from random. OpenAI o3, despite its advanced chain-of-thought reasoning architecture and strong general scientific performance, reaches only ROC-AUC $\approx$ 0.59 at best. These results are striking because both models represent the state of the art in general-purpose AI reasoning, with documented capabilities in mathematics, chemistry knowledge, and multi-step logical inference. Their failure on bioactivity prediction is not a failure of intelligence, it is a failure of the task-fit between reasoning-by-analogy and the empirical, high-dimensional structure-activity landscape that bioactivity prediction demands at the inference level. The frontier LLMs evaluated here have no mechanism for internalising the specific binding-activity relationships of antimalarial targets through a few SMILES-label examples. The gradients produced during fine-tuning provide the mechanism by which these specific dependencies are internalised into model parameters; in-context learning cannot substitute for this process regardless of the base model\textquotesingle s general capability.
More broadly, these results reinforce the finding from LlaSMol \cite{yu2024llasmol} and prior ICL benchmarks \cite{guo2023largelanguagemodel}.

\subsection{Does Biomedical Pretraining Confer a Genuine Advantage? Dissecting TxGemma\textquotesingle s ROC-AUC Lead}\label{does-biomedical-pretraining-confer-a-genuine-advantage-dissecting-txgemmas-roc-auc-lead}
Among fine-tuned models, TxGemma-9B achieves the highest ROC-AUC (0.731 ± 0.005), outperforming both the Random Forest baseline (0.683) and the general-purpose Gemma-2-9B (0.638). 
The within-scale comparison between TxGemma-9B (ROC-AUC: 0.731) and Gemma-2-9B (ROC-AUC: 0.638) models identical in architecture but differing in the domain of their pretraining fine-tune, provides the cleanest experimental window onto this question. The 9.3-percentage-point ROC-AUC advantage of TxGemma-9B over its general-purpose counterpart at equivalent scale is consistent with the hypothesis that biomedical pretraining confers a genuine and transferable advantage for molecular bioactivity classification. We interpret this advantage as arising from TxGemma\textquotesingle s extensive exposure to diverse molecular property prediction tasks during TDC fine-tuning, which likely shapes the model\textquotesingle s internal representations of SMILES tokens towards biologically grounded features, including patterns associated with pharmacophoric activity, target-binding motifs, and ADMET-related structural characteristics.
Interestingly, the same comparison at the 2B scale reveals a much narrower performance gap: TxGemma-2B achieves ROC-AUC 0.695 versus Gemma-2-2B\textquotesingle s 0.706, with Gemma-2-2B marginally superior. The 2B scale comparison complicates the domain-pretraining narrative in a way that warrants explicit acknowledgement. At 9B parameters, TxGemma outperforms Gemma-2 by 9.3 ROC-AUC percentage points, a gap consistent with biomedical pretraining providing a meaningful representational prior. At 2B parameters, this advantage inverts: Gemma-2-2B achieves ROC-AUC 0.706 against TxGemma-2B\textquotesingle s 0.695. We interpret this reversal as a capacity-dependent interaction: at smaller parameter scales, the gradient signal from Malaria-Instruct fine-tuning is sufficient to overcome the head-start conferred by TDC pretraining, whereas at 9B the richer representational capacity of the larger model allows TxGemma\textquotesingle s biomedical priors to be more effectively leveraged during task-specific adaptation. However, these findings require more extensive evaluation in subsequent works.
The biomedical specialisation of TxGemma confers a measurable advantage specifically at larger model scales, consistent with the hypothesis that the benefit of domain pretraining scales with the model\textquotesingle s capacity to utilise the additional representational priors it provides. This interaction between model scale and domain pretraining has important practical implications: research groups with access only to smaller models may find that the additional overhead of deploying specialised therapeutic LLMs yields limited marginal benefit over general-purpose
alternatives at comparable parameter counts.

\subsection{Chemistry-Aware Pretraining and Enrichment: Interpreting LlaSMol\textquotesingle s Superior Enrichment Factor}\label{chemistry-aware-pretraining-and-enrichment-interpreting-llasmols-superior-enrichment-factor}
While TxGemma-9B achieves the highest ROC-AUC, LlaSMol-Mistral attains the best enrichment factor at the 1\% level (EF@1\% $\approx$ 4.99 ± 0.005), substantially exceeding all other models --- including TxGemma-9B (EF $\approx$ 4.23), Random Forest (EF $\approx$ 2.58), and XGBoost (EF $\approx$ 1.79). LlaSMol-Mistral also achieves the highest MCC (0.564 ± 0.026), suggesting that its decision boundary is better calibrated for the class-imbalance conditions characteristic of antimalarial screening data. These results indicate that chemistry-aware pretraining on SMolInstruct encodes structural pharmacophoric features that are particularly well-suited to prospective hit enrichment, even though LlaSMol-Mistral\textquotesingle s global discriminative performance (ROC-AUC: 0.702) is not the highest in the cohort.
A model can achieve high EF@1\% through excellent concentration of actives in the extreme tail of its score distribution, even if its overall ranking is not perfect and LlaSMol-Mistral appears to achieve precisely this: exceptional precision at high confidence thresholds, reflecting its chemistry-aware pretraining\textquotesingle s ability to assign high scores specifically to structurally credible antimalarial scaffolds. We hypothesise that this enrichment advantage arises from LlaSMol\textquotesingle s training on SMolInstruct\textquotesingle s diverse chemistry tasks, which include molecular property prediction, reaction prediction, and molecular name conversion. This multi-task molecular training likely equips LlaSMol with a richer internal vocabulary of substructure-property associations, including the recognition of pharmacophoric features associated with antiplasmodial activity, that is directly leveraged during Malaria-Instruct fine-tuning. By contrast, TxGemma\textquotesingle s TDC pretraining is broader in biological scope (spanning cell lines, proteins, and clinical outcomes) but may be less densely populated with the structural chemistry examples needed for fine-grained pharmacophoric enrichment.
The coexistence of TxGemma-9B\textquotesingle s ROC-AUC lead and LlaSMol-Mistral\textquotesingle s EF@1\% lead is not a contradiction to be resolved but a model-selection signal to be acted upon. For researchers reporting to a benchmark leaderboard or comparing models on global discriminative capacity, TxGemma-9B is the recommended choice. For researchers deploying a model in a prospective VS campaign, where the only compounds that reach experimental assay are those in the top-ranked fraction, LlaSMol-Mistral is the operationally superior model: its EF@1\% of 4.99 concentrates nearly five true actives for every one expected by chance, outperforming TxGemma-9B\textquotesingle s 4.23 by a margin that translates directly into measurable reductions in experimental cost per confirmed hit.
\subsection{The ROC-AUC vs. Enrichment Factor Tension: What Are We Actually Optimising For in Virtual Screening?}\label{the-roc-auc-vs.-enrichment-factor-tension-what-are-we-actually-optimising-for-in-virtual-screening}
The discordance between ROC-AUC and EF@1\% rankings in our results raises a fundamental question about model selection for VS applications: which metric should govern the choice of model for practical deployment? We argue that EF@1\% is the operationally primary metric for VS, and that optimisation for ROC-AUC alone may lead to suboptimal model selection in resource-constrained screening campaigns.
In a practical VS workflow, the output of a computational model is used to rank a chemical library of potentially millions of compounds, from which a small fraction, typically 1\% or less, is selected for experimental synthesis and assay. The productivity of the screening campaign is determined almost entirely by the hit rate in this top-ranked fraction. A model that achieves ROC-AUC 0.731 but EF@1\% of 4.23 (TxGemma-9B) produces a smaller fraction of confirmed actives per experimental unit cost than a model with ROC-AUC 0.702 but EF@1\% of 4.99 (LlaSMol-Mistral), even though its global discriminative performance appears superior. From the perspective of return on investment in experimental screening, LlaSMol-Mistral is the preferred model, a conclusion that would be obscured by relying exclusively on ROC-AUC.
This consideration has direct implications for the design of training objectives in molecular ML. Many existing bioactivity prediction models are trained with cross-entropy loss, which optimises a global classification objective and implicitly targets ROC-AUC-adjacent performance. 

\subsection{Classical Models Remain Competitive Baselines But at What Cost?}\label{classical-models-remain-competitive-baselines-but-at-what-cost}
The Random Forest baseline achieves a ROC-AUC of 0.683 and EF@1\% of 2.58, which while significantly below the performance of the best fine-tuned LLMs, constitutes a strong and highly resource-efficient result. For large-scale VS campaigns screening libraries of tens of millions of compounds, this difference in inference latency is non-trivial: LlaSMol-Mistral would require approximately 93 GPU-days to screen a 100-million-compound library, whereas Random Forest would accomplish the same task in under an hour on a single CPU core.
It should also be noted that the performance advantage of fine-tuned LLMs over classical models, is measured under strict scaffold-dissimilarity conditions. Under random splitting, the gap between classical and LLM-based models is likely to be substantially reduced, as both model classes benefit from the ability to interpolate between structurally similar training and test molecules.

\subsection{Reproducibility and Variance: What the Experimental Design Reveals}\label{reproducibility-and-variance}
The experimental designs employed for fine-tuned and few-shot models are not parallel replication schemes, they are structurally distinct and answer different inferential questions. Interpreting the variance patterns in each case requires that distinction to be made explicit.
For fine-tuned models, all experiments were conducted in duplicate with fully independent Lo-Hi resplitting of the Malaria-Instruct corpus for each replicate. Each duplicate therefore constitutes a genuinely independent experimental unit: a different dissimilarity-enforcing partition, a different training trajectory, and a different sequence of weight updates. The resulting standard deviations capture the
\emph{compound variance} of both partition-induced distributional shift and training stochasticity simultaneously. That the observed standard deviations remain small across all fine-tuned models: TxGemma-9B: ROC-AUC ±0.005; LlaSMol-Mistral: EF@1\% ±0.005, is therefore a materially strong stability claim. It asserts that fine-tuned model performance is consistent regardless of which valid Lo-Hi partition is drawn from the corpus, a property that would not be guaranteed given the structural constraints of dissimilarity splitting and the relatively small size of individual assay-level subsets. This stability simultaneously validates the models and the benchmark: Malaria-Instruct produces evaluation conditions that are reproducible under the most demanding splitting conditions the cheminformatics literature currently prescribes.
For few-shot models, the design is structurally different in a way that strengthens, rather than weakens, the null-result argument. The dataset was independently resplit before constructing each shot-count condition. This means variation observed across shot counts reflects the \emph{joint effect} of shot count and partition change simultaneously, the two sources of variance are intentionally confounded within the shot-count comparison.
The inferential consequences of this design and high variance of closed-source models are further discussed in appendix G and H respectively. 

\subsection{Computational Accessibility and Practical Deployment Considerations}\label{computational-accessibility-and-practical-deployment-considerations}
A defining feature of this study is its focus on resource-constrained computational environments. All fine-tuned models in this study required either a 16GB or 24GB GPU for training and a minimum of 24 GB GPU for inference, corresponding to current mid-range GPU configurations (e.g., NVIDIA A100 40GB, available through Google Colab Pro+). Inference time per molecule ranges from 0.004 s (Gemma-2-2B) to 0.080 s (LlaSMol-Mistral), corresponding to throughputs of approximately 12,500 and 750 molecules per minute, respectively. These figures define the practical boundaries of VS campaign scale for each model: at 750 molecules per minute, LlaSMol-Mistral could screen a 100,000-compound library in approximately 2.2 hours on a single GPU, a feasible timeline for academic VS campaigns.
The scatter plot of inference time per molecule versus ROC-AUC and EF@1\% (Figure 2) reveals an important efficiency frontier: Gemma-2-2B and TxGemma-2B occupy the optimal quadrant of high performance and low inference latency, processing approximately 4--7 ms per molecule while achieving ROC-AUC \textgreater{} 0.695. These models represent the most resource-efficient operating point for VS applications where throughput is a primary constraint. LlaSMol-Mistral, despite its superior enrichment performance, requires approximately 20-fold more inference time, a trade-off that is acceptable for smaller library screening but may be prohibitive at scale. A significant caveat is that the inference time figures reported in Table 5 should be interpreted in the context of the inference framework employed. All reported per-molecule inference times were obtained using vLLM \cite{kwon2023vllm} as inference under standard PyTorch, the default framework for model deployment, is substantially slower by one to two orders of magnitude for all LLMs evaluated with extended analysis of why provided in appendix H.
The QLora parameter-efficient fine-tuning approach (4-bit quantization, rank 8) adopted in this study is instrumental in achieving these resource efficiency figures. By reducing the effective memory footprint of 7--9B parameter models to the 24 GB GPU RAM regime, QLora enables the fine-tuning and deployment of models that would otherwise require 40--80 GB GPU configurations, extending their accessibility to a substantially broader research community. These results demonstrate that instruction-tuned open-source LLMs fine-tuned via QLoRA represent a practically achievable, resource-efficient alternative to classical cheminformatics pipelines for antimalarial VS in low-resource settings.
This characterisation directly addresses the GPU-constrained reality of researchers in malaria-endemic settings, for whom free-tier platforms (Colab, Kaggle) are often the only available compute.
\subsection{Limitations}\label{limitations}
Several limitations of this study should be acknowledged. First, Malaria-Instruct is derived exclusively from the ChEMBL Legacy Malaria corpus, which while comprehensive, captures only chemotypes that have historically been screened against Plasmodium, leaving dark chemical space, novel structural classes not previously investigated, entirely unexplored. The generalisation performance reported here therefore characterises structural novelty within the known antimalarial chemical universe, not the full scope of potential drug-like space.
Second, all molecular representations in this study are SMILES-based, treating molecules as linear strings without explicit encoding of three-dimensional conformational information, stereochemical preferences, or target-binding geometries. Multi-modal representations that integrate SMILES with 3D structural features, molecular graphs, or protein pocket descriptors may offer substantially improved predictive performance, particularly for target-specific activity prediction where the binding site geometry is the primary determinant of activity.
Third, the evaluation presented here is exclusively in silico: no experimental validation of model-predicted hits has been conducted. The practical utility of these models for prospective VS ultimately rests on the chemical and biological validity of their top-ranked predictions, which requires wet-lab confirmation. Prospective experimental validation of model-predicted antimalarial candidates represents the essential next step from this work.
Lastly, the dataset reflects heterogeneity in ChEMBL assay provenance: different assays employ different Plasmodium strains (P. falciparum 3D7, Dd2, K1), measurement timepoints (48h vs. 96h), and readout modalities. While assay-level splitting and contextualisation partially address this heterogeneity, residual inter-assay variability may introduce label noise that differentially affects model classes.

\section{Conclusion and Future Directions}
The central empirical finding of this work is unambiguous: domain-specific fine-tuning is a categorical prerequisite for reliable antimalarial bioactivity prediction with LLMs. The 23-percentage-point ROC-AUC collapse observed in TxGemma-9B from 0.731 under fine-tuning to 0.499 under few-shot in-context learning best describes this with the pattern consistent across all model classes, including frontier proprietary models. Among fine-tuned models, biomedical and chemistry-aware pretraining both confer measurable advantages, but in ways that are metric-dependent and scale-sensitive. TxGemma-9B achieves the highest global discriminative performance (ROC-AUC: 0.731 ± 0.005), with its advantage over general-purpose Gemma-2-9B concentrated at the 9B parameter scale, suggesting a capacity-dependent interaction between domain pretraining and task-specific adaptation that diminishes at smaller model sizes while LlaSMol-Mistral, achieves the highest enrichment factor (EF@1\%: 4.99 ± 0.005) and MCC (0.564 ± 0.026), outperforming all models on the metrics most directly relevant to prospective screening campaign productivity.
Classical machine learning baselines also remain competitive reference points with significant throughput advantage over fine-tuned LLMs spans three to four orders of magnitude, and for VS campaigns operating at library scales of tens of millions of compounds, suggesting a tiered strategy combining classical pre-filtering with LLM re-ranking of shortlisted candidates may represent the optimal resource allocation. The inference time and GPU memory characterisation provided in this study, offers the quantitative foundation needed to make these deployment decisions on an evidence-based rather than intuitive basis.
Several promising directions also emerge from this work. However, the most important is the prospective experimental validation of top-ranked model predictions, synthesising and screening the highest-confidence predicted actives from diverse structural clusters, represents the ultimate test of VS model utility. Such validation would provide both scientific evidence of practical impact and training data for subsequent model improvement cycles, closing the loop between computational prediction and experimental drug discovery.

\bibliographystyle{named}
\bibliography{ijcai26}

\section*{Appendix}
\subsection*{A. ChEMBL Dataset}
ChEMBL is a manually curated, large-scale open-access bioactivity database that integrates compound-target interaction data extracted from the primary medicinal chemistry literature, depositor submissions, and partner databases \cite{mendez2019chembl}\cite{zdrazil2024chembl}. At present, ChEMBL contains bioactivity data for over two million distinct compounds, making it the most comprehensive public resource for QSAR modelling and computational drug discovery. Its ChEMBL Legacy Malaria dataset, compiled in partnership with the Medicines for Malaria Venture and aggregating decades of antimalarial screening data, constitutes one of the richest freely available sources of antimalarial bioactivity data, spanning multiple assay formats, Plasmodium strains, and time periods.
\subsection*{B. Instruction Tuning and Dataset Curation for Scientific LLMs}
The instruction tuning paradigm, in which language models are fine-tuned on curated collections of (instruction, response) pairs to improve instruction-following fidelity and task generalisation, has fundamentally transformed the practical utility of large language models \cite{wei2021finetuned}\cite{ouyang2022training}. In the domain of scientific LLMs, instruction tuning has been applied to produce models capable of responding to molecular queries in natural language, with the instruction format serving as a compositional scaffold that enables multi-task learning across diverse property prediction objectives.
Prior benchmark datasets for molecular instruction tuning include Mol-Instructions \cite{fang2023molinstructions} and SMolInstruct \cite{yu2024llasmol}. Mol-Instructions spans molecule-text translation, property prediction, and molecular design tasks, while SMolInstruct extends coverage to 14 distinct chemistry tasks with rigorous quality control. Models trained on InstructMol \cite{cao2024instructmol} integrate molecular graph representations with language instructions, leveraging structural inductive biases that pure SMILES-text models lack. The Tx-LLM and TxGemma data preparation pipeline \cite{zambrano2024txllm}, \cite{wang2025txgemma} further extends instruction tuning to multi-modal therapeutic tasks that integrate molecular, protein, cell line, and disease ontology information within a single prompt framework.
\subsection*{C. Biomedical Domain Specific Pretrained Models}
Tx-LLM \cite{zambrano2024txllm}, the direct predecessor to TxGemma, was a generalist therapeutic LLM fine-tuned from PaLM-2 on 709 datasets targeting 66 tasks spanning the drug discovery pipeline. Tx-LLM demonstrated that a single set of model weights could simultaneously encode knowledge about small molecules, proteins, nucleic acids, cell lines, and diseases, achieving near-state-of-the-art performance on 43 of 66 Therapeutics Data Commons (TDC) tasks \cite{huang2021tdc}.
TxGemma \cite{wang2025txgemma} improves on Tx-LLM by adopting the Gemma-2 architecture and retraining on the full TDC benchmark suite of 66 therapeutic tasks using 7 million curated examples. TxGemma-Predict, the predictive variant is available in 2B, 9B, and 27B parameter configurations, all of which have been benchmarked against specialist models on the TDC suite. Notably, TxGemma-9B-Predict has been validated to improve over Tx-LLM on 45 of 66 TDC tasks and to match or exceed best-in-class specialist model performance on 50 tasks 

\subsection*{D. Evaluation considerations}
The decision of which evaluation metric is chosen shapes the conclusions drawn from VS model benchmarks. Accuracy, the fraction of correctly classified instances, is a misleading primary metric under class imbalance: a model that always predicts the majority class (inactive) can achieve greater than 90\% accuracy while providing no practical utility for VS. ROC-AUC provides a global measure of discriminative capacity that is insensitive to class imbalance and corresponds to the probability that a randomly drawn active is ranked above a randomly drawn inactive, making it a standard and interpretable benchmark metric. Matthews Correlation Coefficient (MCC) offers a balanced single-number summary that incorporates all four cells of the confusion matrix and is considered more informative than F1 under severe class imbalance \cite{chicco2020advantages}.
\subsection*{E. Need for Strong Baseline}
Praski \textit{et al.} (2026) \nocite{praski2026benchmarking} cautions against the assumption that architectural complexity necessarily translates to improved generalisation. In a systematic comparison of deep learning approaches to fingerprint-based baselines, they found that complex deep learning approaches offer negligible or non-significant improvements to fingerprint-based baselines in many bioactivity prediction settings, particularly under realistic data-splitting conditions. This necessitated our inclusion of Random Forest and XGBoost as primary classical baselines, not merely as weak foils, but as genuinely competitive reference points that LLMs must measurably surpass to justify their substantially greater computational cost.

\subsection*{F. Training Parameters}
All LLMs were fine-tuned using QLoRA \cite{dettmers2023qlora} with 4-bit quantisation, LoRA rank 8, and the following target modules: q\_proj, o\_proj, k\_proj, v\_proj, gate\_proj, up\_proj, and down\_proj. Training was conducted for one epoch over the full training partition. Optimisation used the paged AdamW 8-bit optimiser with a learning rate of $2\times10^{-4}$, weight decay of 0.001, and a warmup of 2 steps. Batch configuration was set to a train batch size of 2 with gradient  accumulation of 2, or a train batch size of 4 with gradient accumulation of 1, depending on available GPU memory.

\subsection*{G. Inference Consequence of Few-shots resplitting}
The confounding impart of the dataset being independently resplit before constructing each shot-count condition is analytically. While the variation observed across shot counts reflects the \emph{joint effect} of shot count and partition change simultaneously, it answers the question that if in-context learning were genuinely capable of extracting structural bioactivity signals, performance should either be consistently good or improve monotonically as shot count increases \emph{even across different partitions}, because each additional SMILES-label example should convey incrementally useful structure-activity information irrespective of the evaluation neighbourhood. The absence of any such monotonic improvement and in several cases outright performance degradation from 3-shot to 5-shot cannot be attributed to a fixed partition artefact, because the partition changes with each condition. The near-random ICL performance is therefore robust to both variables simultaneously, substantially strengthening the conclusion that the failure is intrinsic to the task-model mismatch rather than an accident of a particular structural partition.

\subsection*{H. Closed Source Few-Shot Variance}
The high variance observed in closed-source model performance under few-shot conditions carries a distinct and more pointed interpretation. Gemini 2.5 achieves EF@1\% ranging from 2.34 to 3.79 across duplicate evaluations at 4-shot (±2.499), despite the shot count and partition being held constant within each duplicate pair with only the identity of the specific few-shot examples varying. This sensitivity to example identity is the behavioural signature of a model without genuine task grounding: a model that has internalised the structural determinants of antimalarial bioactivity would produce consistent rankings regardless of which specific SMILES-label pairs appear in the prompt. A model responding sensitively to the idiosyncratic features of individual example molecules is instead retrieving surface-level patterns from its prompt context. The high variance in closed-source ICL performance is therefore not methodological noise to be controlled away, it is itself evidence of the fundamental instability of in-context bioactivity prediction, and should be interpreted as such.

\subsection*{I. Inference Engine for LLMs: vLLM vs PyTorch Native}
The source of this gap is between pytorch native implementation and vLLM is architectural rather than hardware-dependent. Autoregressive generation in standard PyTorch allocates GPU memory for the key-value cache on a per-sequence basis without batching or memory reuse across requests, producing significant GPU idle time between token generation steps even for short outputs. Because activity label prediction requires generating only one or two tokens per molecule, the ratio of cache initialisation overhead to productive computation is particularly unfavourable, amplifying the inefficiency relative to longer-form generation tasks where the fixed overhead is amortised across many tokens. vLLM\textquotesingle s PagedAttention mechanism eliminates this fragmentation by managing the KV cache in fixed non-contiguous memory blocks, enabling continuous batching across variable-length requests and achieving near-complete GPU memory utilisation. The implication for practitioners is significant: the inference times reported here are achievable only with vLLM or a comparable optimised inference framework.
Researchers deploying fine-tuned LLMs for VS using standard PyTorch should anticipate inference throughputs one to two orders of magnitude below those reported in Table 5, which would render library-scale screening with LlaSMol-Mistral or TxGemma-9B computationally infeasible without dedicated inference optimisation. Reporting inference times without specifying the inference framework, a common omission in the LLM benchmarking literature, therefore produces figures that are not reproducible in standard deployment contexts and may substantially overstate the practical accessibility of LLM-based VS.

\end{document}